\documentclass[12pt]{article}

\usepackage{amsmath,amssymb,array,amsfonts}
\usepackage[dvipdfmx]{graphicx}
\usepackage{comment,color}
\usepackage{cite}

\newcommand{\bs}{\boldsymbol}
\newcommand{\ba}{\begin{eqnarray}}
\newcommand{\ea}{\end{eqnarray}}

\allowdisplaybreaks

\def\ncm{\newcommand}
\def\e {{\rm e}}
\def\s {{\rm s}}
\def\H {{\rm H}}
\def\B {{\rm B}}
\def\SM{{\rm SM}}
\def\T {{\rm T}}

\def\Tr{{\rm Tr}}
\def\d{{\rm d}}

\def\dis{\displaystyle}

\def\B{{\rm B}}
\def\nt{\notag}

\ncm{\sls}[1]{{\ooalign{\hfil/\hfil\crcr$#1$}}}

\makeatletter
    
    \@addtoreset{equation}{section}
\makeatother

\begin{document}
\setlength{\baselineskip}{18pt}
\begin{titlepage}

\begin{flushright}
OCU-PHYS 417
\end{flushright}
\vspace{1.0cm}
\begin{center}
{\Large\bf Deviation of Yukawa Coupling \\
\vspace*{3mm}
in Gauge-Higgs Unification} 
\end{center}
\vspace{25mm}

\begin{center}
{\large
Yuki Adachi 
and 
Nobuhito Maru$^{*}$
}
\end{center}
\vspace{1cm}
\centerline{{\it
Department of Sciences, Matsue College of Technology,
Matsue 690-8518, Japan.}}

\centerline{{\it
$^{*}$
Department of Mathematics and Physics, Osaka City University, Osaka 558-8585, Japan.
}}
%
%
\vspace{2cm}
\centerline{\large\bf Abstract}
\vspace{0.5cm}
We study the deviation of yukawa coupling in the gauge-Higgs unification scenario from the Standard Model one. 
Taking into account the brane mass terms necessary for generating the flavor mixing and removing the exotic massless fermions, 
 we derive an analytic formula determining the KK mass spectrum and yukawa coupling. 
Applying the obtained results to the tau and bottom yukawa couplings, 
 we numerically calculate the ratio of the yukawa couplings in the gauge-Higgs unification and in the Standard Model. 
\end{titlepage}




\newpage
\section{Introduction}

Although a Higgs boson was discovered at the CERN Large Hadron Collider (LHC) experiment \cite{Higgs}, 
 the couplings of the Higgs boson to the Standard Model (SM) fields and the self-couplings of Higgs boson have not been precisely measured.  
It is therefore still an unsettled issue whether the Higgs boson is the SM one or that of physics beyond the SM. 
Physics beyond the SM is expected to exist by several reasons such as the hierarchy problem.

Gauge-Higgs unification (GHU) \cite{GH} is one of the attractive scenarios beyond the SM, 
 which provides a possible solution to the hierarchy problem without supersymmetry \cite{HIL}. 
In this scenario, 
 the SM Higgs boson and the gauge fields are unified into the higher dimensional gauge fields, 
 {\em i.e.} Higgs boson is identified with extra spatial components of higher dimensional gauge fields.
A remarkable fact is that the quantum correction to Higgs mass (and potential) is UV-finite
 and calculable due to the higher dimensional gauge symmetry 
 though the theory is the non-renormalizable.
The finiteness of the Higgs mass has been studied by explicit diagrammatic calculations 
 and verified in models with various types of compactification 
 at one-loop level 
 \cite{ABQ}
 and even at the two loop level \cite{MY}. 
The finiteness of other physical observables such as $S$ and $T$ parameters \cite{LM}, 
 Higgs couplings to digluons, diphotons \cite{Maru}, Muon $g-2$ and the EDM of neutron \cite{ALM} 
 have been investigated by the present authors or one of them.

The fact that the Higgs boson is a part of gauge fields implies that
 Higgs interactions are governed by gauge principle
 and may provide specific predictions in LHC physics. 
From this viewpoint, the diphoton and $Z\gamma$ decay of the SM Higgs boson produced via the gluon fusion  
 in the framework of gauge-Higgs unification was studied and remarkable predictions were obtained \cite{MO}. 
In order to explain experimental results of diphoton decay and 126 GeV Higgs boson mass, 
 some extra matters are required and they may predict a possible dark matter candidate. 
It has been also shown that the $Z\gamma$ decay is not affected at one-loop level, 
 which is a distinctive prediction uncommon in other models of physics beyond the SM. 
Thus, $Z\gamma$ decay is considered to be a good probe of GHU.

In this paper,
 we focus on the fermion coupling of Higgs boson in the GHU scenario, {\em i.e.} Yukawa coupling, 
 whose measurement in future would be very important to clarify the origin of the Higgs field. 
In GHU scenario, yukawa coupling generically deviates from the SM one 
 as a consequence of the Higgs boson as a gauge field. 
Let us parametrize the fermion mass term as
\ba
m(v) \bar{\psi}\psi,
\ea 
where $m(v)$ is a mass function of the vacuum expectation value (VEV) of Higgs field. 
Physical Higgs coupling to fermions are obtained by expanding the Higgs field around its VEV $v$, 
\ba
m(v+h) \bar{\psi} \psi = m(v) \bar{\psi} \psi + \frac{\d m(v)}{\d v} h \bar{\psi} \psi + \cdots,
\ea 
where $h$ is a physical Higgs field 
 and the second term is its coupling to fermions of our interest in this paper.  
Its coefficient $f \equiv \frac{\d m(v)}{\d v}$ is yukawa coupling.   
$\cdots$ implies higher order terms in $h$ which are irrelevant throughout this paper. 
In GHU, the Higgs field is a zero mode of gauge field $A_y^{(0)}$ for five dimensional case. 
If the fifth dimension $y$ is compactified on a circle $S^1$ with a radius $R$, 
 a constant $A_y^{(0)}$ cannot be removed by the gauge transformation 
 and has a physical meaning of Aharanov-Bohm (AB) phase or Wilson-loop as
\ba
W = P {\rm exp} \left[ i \frac{g}{2} \oint_{S^1} A_y \d y \right] 
= {\rm exp} \left[ ig_4 \pi R A_y^{(0)} \right],
\label{Wilson}
\ea 
where $g, g_4$ are 5D and 4D gauge couplings, respectively. 
An important point of eq.~(\ref{Wilson}) is that $W$ is periodic 
 with respect to $A_y^{(0)}$ under $A_y^{(0)} \to A_y^{(0)}+ 2/(g_4 R)$. 
This fact is one of the characteristic features of GHU 
 and the physical observables are expected to have the periodicity in the Higgs field. 
Actually, it has been already reported in \cite{GHYukawaflat} in the flat extra dimensional case 
 and \cite{GHYukawaWarped} in the warped extra dimensional case 
 that Higgs coupling to the fermions have such a periodicity. 
In \cite{GHYukawaflat}, the ratio of yukawa coupling of GHU and the SM one is derived as 
\ba
\frac{f_{{\rm GHU}}}{f_{{\rm SM}}} \simeq \frac{g_4}{2}v \pi R \cot \left( \frac{g_4}{2} v\pi R \right),
\ea  
which is quite distinctive from the other models beyond the SM. 
In particular, as was pointed out in \cite{GHYukawaflat, GHYukawaWarped},
the yukawa coupling vanishes at $v=1/(g_4 R)$ due to the periodicity. 

In the minimal supersymmetric standard model (MSSM) case, 
 the corresponding ratio is  known to be
\ba
\frac{f_{{\rm MSSM}}}{f_{{\rm SM}}} = 
\left\{
\begin{array}{l}
\frac{\cos \alpha}{\sin \beta}~({\rm up-type~quarks}), \\
- \frac{\sin \alpha}{\cos \beta}~({\rm down-type~quarks~and~charged~leptons}), \\
\end{array}
\right. 
\ea
where $\alpha$ is a mixing angle of two Higgs doublets in the MSSM 
 and $\beta$ is defined as $\tan \beta = \langle H_u \rangle/\langle H_d \rangle$. 
As for the UED models, the yukawa coupling is the same as the SM one. 
Thus, it is a very important issue for the new physics search to measure the yukawa couplings precisely at LHC and ILC.  

We study the deviations of Higgs coupling to fermions from the SM predictions 
 by using a five dimensional GHU model of $SU(3)$ gauge theory. 
Unlike the analysis where only the bulk Lagrangian was considered \cite{GHYukawaflat}, 
 we take into account the effects from the brane mass terms necessary in a more realistic model 
 for generating the flavor mixing as clarified in \cite{flavorGHU} 
 and removing the exotic massless fermions absent in the SM. 
The brane mass terms change the boundary conditions of mode equations, 
 which give the formula to determine Kaluza-Klein (KK) mass spectrum $m_n(v)$. 
Therefore, it is crucial for the study of the deviation of Yukawa coupling in a realistic model 
 to incorporate the appropriate brane mass terms. 
Solving the mode equation with the correct boundary conditions, 
 we will derive analytic formulas of determining KK mass spectrum 
 and the ratio of Yukawa coupling in GHU and the SM one. 
As an illustration, we numerically calculate its ratio for the tau and the bottom yukawa couplings, 
 which are expected to be more promising detectable couplings in the quark and lepton sector 
 comparing to those of other lighter fermions.

This paper is organized as follows. 
In section 2, we introduce our model.  
We elaborate the equations of motion and the corresponding boundary conditions. 
Analytic formulas determining KK mass spectrum and the expression of yukawa coupling are derived in section 3. 
Numerical calculations for the ratio of the tau  and the bottom yukawa couplings in GHU and the SM ones 
 are also performed as an application. 
Section 4 is devoted to summary. 
In appendix A, the derivation of the equations of motion is described in detail. 
The validity of the analytic formula obtained in this paper is checked 
 by taking various limits of parameters in appendix B.

\section{The Model}
In this section, we introduce our model. 
We consider an $SU(3)$ gauge theory in a five dimensional space-time 
 where an extra dimension is compactified on an orbifold $S^1/Z_2$.
The Lagrangian consists of two parts; 
\begin{equation}
	\mathcal L = \mathcal L_{\rm Bulk} + \mathcal L_{\rm Brane}.
\end{equation}
One is the Lagrangian in the bulk where the extra dimensions spread  
and the other is that on the brane located at fixed points $y=0, \pi R$.
The bulk Lagrangian is 
\begin{align}
	\mathcal L_{\rm Bulk}
	=&
	\bar{\psi} ({\bf 3}) (i \not \!\!\!\!\ D_{\bf 3}- M \epsilon(y)) \psi({\bf 3})
	+ \bar{\psi}({\bf 6}^\ast) (i \not \!\!\!\!\ D_{\bf 6^\ast}- M \epsilon(y)) \psi({\bf 6^\ast})
	-\frac12 \Tr F_{MN}F^{MN},
\end{align}
where the $\bf 3$ and the $\bf 6$ stands for the three and six dimensional representations of $SU(3)$, 
the covariant derivative is given by the $SU(3)$ gauge field $\not \!\!\!\!\ D=\Gamma^M(\partial_M -ig A_M^aT^a)$ 
with the appropriate generators $T^a$ for the corresponding representation, 
and the field strength of $SU(3)$ is denoted by $F_{MN}$. 
The components of $\bf 3$ and $\bf 6^\ast$ are 
\begin{equation}
\psi({\bf 3})=
\begin{pmatrix}
	\nu\\\tau\\ \tau_1
\end{pmatrix}
,
\psi({\bf  6^\ast})=
\begin{pmatrix}
\Sigma_\downarrow & -\frac1{\sqrt{2}} \tau_3 & \frac1{\sqrt{2}} \tau'\\
-\frac1{\sqrt{2}}\tau_3 & \Sigma_\uparrow & \frac1{\sqrt{2}} \nu\\
\frac1{\sqrt{2}}\tau' & \frac1{\sqrt{2}}\nu & \frac1{\sqrt{2}} \nu_\s
\end{pmatrix}
.
\end{equation}
The $\tau$ field belongs to different isospin but has the same electric charge as $\tau$ lepton,  
 so that  they will mix each other by the VEV of $A_y$. 
This matter contents are common in quarks and leptons except for the top quark.
The fermion mass is at most of order W-boson mass and the top quark has a mass around twice of the W-boson mass, 
 so it has to be embedded in higher dimensional representation such as $\overline{\bf 15}$ for example 
 to obtain the enhancement factor ``2".

Since the fifth dimension $y$ is compactified on the circle $S^1$, 
the periodic boundary condition is imposed for the fields $\phi_i$; 
\begin{equation}
	\phi_i(y) = \phi_i(y+2\pi R).
\end{equation}
To have the chiral theory, the $Z_2$ symmetry is imposed on the fermions.
By assigning even eigenvalue of $Z_2$ to the right-handed singlet and left-handed doublets for the SM fermions,
their zero modes remains massless. 
For the singlet lepton $\psi$, the $Z_2$ eigenvalues  are assigned as
\begin{equation}
	\psi_L(+y) = -\psi_L(-y),
	\psi_R(+y) = \psi_R(-y).
	\label{}
\end{equation}
For the lepton doublets $\chi$, they becomes
\begin{equation}
	\chi_L(+y) = \chi_L(-y),
	\chi_R(+y) = -\chi_R(-y),
	\label{}
\end{equation}
where the chiral projection operators are defined as $L=\frac{1-\gamma^5}{2}$ and $R=\frac{1+\gamma^5}{2}$.
For other cases, the $Z_2$ parities are given by
\begin{equation}
	\psi(+y) = \gamma^5\psi(-y),
	\chi(+y) = -\gamma^5 \chi(-y).
	\label{}
\end{equation}
We note that the $SU(3)$ gauge symmetry is simultaneously broken to $SU(2) \times U(1)_Y$ 
 and the SM Higgs doublet is realized in $A_y^{(0)}$ by the appropriate $Z_2$ parity assignment.

In general,  we have extra massless fermions which are  not included in the SM. 
In particular, the two massless $SU(2)$ doublets appear per generation 
 since up and down sector fermions should be embedded in different representations. 
One of the linear combination of them corresponds to the SM $SU(2)$ doublet, 
 but the other orthogonal one should be removed from the low-energy effective theory. 
A possible way is that they couple with the brane-localized fermions ($\tau_\B'$ and $\tau_\B''$)
 and become massive through the Dirac mass terms.
\begin{equation}
	\mathcal L_\text{Brane}
	=
	\sqrt{\pi R}\bar \tau_\B' M_\B\tau_\H \delta(y) 
	+\sqrt{\pi R}M_\B\bar \tau_\B'' \tau_3\delta(y)
	+{\rm h.c.},
	\label{}
\end{equation}
where the $\tau_\H$ is the massive tau leptons orthogonal to the massless tau lepton $\tau_\SM$.
They are  mixing states of the $\tau$ and $\tau'$ which are defined by 
\begin{equation}
	\begin{pmatrix}\tau_1\\\tau\\ \tau'\\ \tau_3\end{pmatrix}
	=
	\begin{pmatrix}
	1 & 0&0&0\\
	0&\cos \theta&\sin\theta&0\\
	0&-\sin\theta&\cos\theta&0\\
	0&0&0&1
	\end{pmatrix}
	\begin{pmatrix}\tau_1\\\tau_\SM\\ \tau_\H\\ \tau_3\end{pmatrix}
	\equiv
	U
	\begin{pmatrix}\tau_1\\\tau_\SM\\ \tau_\H\\ \tau_3\end{pmatrix}
	.
	\label{}
\end{equation}
These mixings play an important role to produce the flavor mixings \cite{flavorGHU}.

To achieve our purposes, we concentrate on the tau leptons and the equation of motion (EOM) 
 derived from the lagrangian as follows;
\begin{align}
\label{EOM_tau}
\left[
 i \partial_\mu\gamma^\mu -\partial_y\gamma^5
 +i\frac{g_4}{2}v \gamma^5 \Sigma_1
 -M\epsilon(y) 
\right]
{\bs \tau}
 =& -\sqrt{\pi R} M_\B  \begin{pmatrix}0\\0\\ \tau_\B'\\\tau_\B''\end{pmatrix} \delta (y),
 \\
 i\partial_\mu\gamma^\mu \tau_\B'\delta(y) =& - \sqrt{\pi R}M_\B\tau_\H \delta(y),
 \\
 i\partial_\mu\gamma^\mu \tau_\B''\delta(y) =& -\sqrt{ \pi R}M_\B\tau_3 \delta(y),
\end{align}
where $\bs \tau =(\tau_1, \tau_\SM, \tau_\H, \tau_3)^{\T}$.
$\Sigma_1$ is defined by 
\begin{equation}
	\Sigma_1 =
	U^\dag
	\begin{pmatrix}
		\sigma_1 & 0 \\ 0 & \sqrt2 \sigma_1 
	\end{pmatrix}
	U,
\end{equation}
where $U$ connects the basis between $(\tau,\tau')$ and $(\tau_\H,\tau_\SM)$ as we mentioned before. 
$\sigma_1$ is a Pauli matrix. 
The factor $\sqrt{2}$ comes from the group theoretical factor of the representation of ${\bf 6^*}$ (two-rank symmetric tensor).  

Defining $\hat {\bs \tau}$ to eliminate $\Sigma_1$ as
\begin{equation}
	\bs\tau = \exp\left[i\frac{g_4}{2}v\Sigma_1 y\right] \hat {\bs \tau}, 
\end{equation}
then we have
\begin{align}
\label{EOM_tau2}
\left[
 i\partial_\mu\gamma^\mu -\partial_y\gamma^5
 -M\epsilon(y) 
\right]
\hat {\bs \tau}
 =&-\sqrt{\pi R} M_\B  \begin{pmatrix}0\\0\\ \tau_\B'\\\tau_\B''\end{pmatrix}\delta (y),
 \\
 i\partial_\mu\gamma^\mu \tau_\B'\delta(y) =& -\sqrt{ \pi R}M_\B\hat\tau_\H \delta(y),
 \\
 i\partial_\mu\gamma^\mu \tau_\B''\delta(y) =& -\sqrt{ \pi R}M_\B\hat\tau_3 \delta(y).
\end{align}
The boundary conditions (B.C.s) of the $\hat {\bs \tau}$ at $y=0$ is same as before, 
 but it changes at $y=\pi R$ because of the Wilson line phases.
We summarize the $Z_2$ conditions on the $\hat \tau $ at the origin $y=0$
	\begin{equation}
	\hat{\bs \tau}
	(-y)
	=
	P\gamma_5 
	\hat{\bs \tau}
	(y),
	\label{}
\end{equation}
where $P=\text{diag}(-,+,+,-)$
 and the periodic B.C. at $y=|\pi R|$ with respect to $S^1$ 
\begin{equation}
	\label{periodicBC}
	\left.
	\exp\left[i\frac{g_4}{2}v\Sigma_1 y\right]
	\hat{\bs \tau}
	\right|_{y=\pi R}
	=
	\left.  
	\exp\left[i\frac{g_4}{2}v\Sigma_1 y\right]
	\hat{\bs \tau}
	\right|_{y=-\pi R}.
\end{equation}
In other words, 
these periodicities are rewritten in terms of the parities at the $y= \pi R$,
namely,
\begin{equation}
	\left.\left[\e^{i\frac{gv}{2}\Sigma_1 y}\hat {\bs\tau}(y)\right]_{\rm odd}\right|_{y=\pi R}=0 
\end{equation}
where $[\dots]_{\rm odd/even}$ stands for extracting odd/even function of $y$. 
The conditions of the derivative $\partial_y\hat {\bs \tau}$  are obtained 
 by integrating the EOM around $y=\pi R$.   
 Then we have the following two conditions from (\ref{periodicBC});
\begin{equation}
	\label{summary_BC}
	\begin{cases}
&\left.\left[\e^{i\frac{gv}{2}\Sigma_1 y}\hat \tau(y)\right]_{\rm odd}\right|_{y=\pi R}=0,
\\[12pt]
&
\left.\left[\e^{i\frac{gv}{2}\Sigma_1 y}\gamma_5\partial_y\hat {\bs \tau}(y)\right]_{\rm odd}\right|_{y=\pi R}
=
-M\left.\left[\e^{i\frac{gv}{2}\Sigma_1 y}\hat {\bs \tau}(y)\right]_{\rm even}\right|_{y=\pi R}.
\end{cases}
\end{equation}

A few comments are listed.
We omit the strong interaction through this paper 
 since our purpose is to investigate the effects of the flavor mixing and brane mass term
 on the deviation of yukawa coupling from the SM one 
 and the strong interaction does not affect the deviation. 

The Weinberg's angle 
 in this model is not consistent with the observed one, 
 which is obtained by introducing an extra $U(1)'$ gauge group or the brane localized kinetic terms. 
This does not affect the deviation of yukawa couplings originated from an $SU(2)$ gauge coupling 
 and we can safely ignore them in this paper.
By adjusting these extra $U(1)'$ charge, the hypercharge can be changed as we like. 
Therefore, the $\tau$ lepton and the $b$ quark can be assigned to the same representations ${\bf 3}$ and ${\bf 6^*}$ 
 by taking different hypercharges \cite{SSS}. 

\section{Deviation of Yukawa Coupling in GHU from the SM one}
In this section, 
 we discuss the deviation of yukawa coupling in GHU from the SM one. 
First of all, we derive the analytic formula determining the KK mass spectrum from the boundary conditions of fermions, 
the continuous conditions at $y=|\pi R|$ and the $Z_2$ condition.
Next we obtain the yukawa coupling in GHU through the analytic formula 
 by differentiating the KK fermion mass $m(v)$ with respect to the VEV $v$. 
As a phenomenological application, 
 numerical calculations for the ratio of the tau and the bottom yukawa couplings in GHU and the SM ones 
 are performed. 
\subsection{Analytic formula determining KK mass spectrum}
To begin with, we expand the $\hat \tau$ in terms of the mode functions as follows;
\begin{align}
	\label{mode_expansion}
	\hat{\bs \tau}(x,y)
=
\sum_{n=0}^{\infty}
\left[
	\begin{matrix}
	\hat\tau_{1L}^{(n)}(x) f_{1L}^{(n)}(y) + \hat\tau_{1R}^{(n)}(x) f_{1R}^{(n)}(y)
	\\
	\hat\tau_{\SM L}^{(n)}(x) f_{\SM L}^{(n)}(y) + \hat\tau_{\SM R}^{(n)}(x) f_{\SM R}^{(n)}(y)
	\\
	\hat\tau_{\H L}^{(n)}(x) f_{\H L}^{(n)}(y) + \hat\tau_{\H R}^{(n)}(x) f_{\H R}^{(n)}(y)
	\\
	\hat\tau_{3L}^{(n)}(x) f_{3L}^{(n)}(y) + \hat\tau_{3R}^{(n)}(x) f_{3R}^{(n)}(y)
	\end{matrix}
\right].
\end{align}
Hereafter, we omit the index $n$ in the mode functions for notational simplicity. 
Then the eigen equations of the mode functions of $\hat\tau_1$ is obtained as
\begin{align}
	\label{eom_13}
\begin{cases}
	m_{n} f_{1L}+(\partial_y-M\epsilon(y))f_{1R}=0,
	\\
	m_{n} f_{1R}+(-\partial_y-M\epsilon(y))f_{1L}=0,
\end{cases}
\end{align}
where $m_n$ is a KK mass eigenvalue.
The mode functions of $\hat\tau_{\SM}$ obey the same eigen equations.
The eigen equations for mode functions of $\hat \tau_\H,\hat \tau_3$ are given 
 by using the integration by parts of  the delta function $\int \d y f(y)\partial_y \delta(y)=-\int \d y \partial_y f(y) \delta(y)$. 
\begin{align}
\begin{cases}
	&[m_n^2-\partial_y^2-2M(\delta(y)-\delta(y-\pi R))+M^2]f_{\H L}=-\pi RM_\B^2f_{\H L}\delta(y),
\\
&[m_n^2-\partial_y^2+2M(\delta(y)-\delta(y-\pi R))+M^2]f_{\H R}=-\pi RM_\B^2f_{\H R}\delta(y),
\end{cases}
\end{align}
for $\hat{\tau}_H$, and the mode functions of $\hat \tau_{3}$ obeys the same eigen 
equations.\footnote{The derivation of these mode equations are described in Appendix \ref{Appendix1}.}

The eigen equations are immediately solved by respecting the $Z_2$ parties at the origin (\ref{summary_BC}) as
\begin{equation}
	\label{mode_function}
\begin{cases}
	f_{1L} \propto \sin ( \sqrt{m_n^2-M^2} y),
	&
	f_{1R} \propto \cos ( \sqrt{m_n^2-M^2} |y|+\alpha_1),
	\\
	f_{\SM L}\propto \cos (\sqrt{m_n^2-M^2} |y|+\alpha_{\SM}),
	&
	f_{\SM R}\propto \sin (\sqrt{ m_n^2-M^2}y),
\\
	f_{\H L}\propto \cos ( \sqrt{ m_n^2-M^2}|y|+\alpha_\H),
	&
	f_{\H R}\propto \sin( \sqrt{ m_n^2-M^2}y),
\\
	f_{3L}\propto \sin (\sqrt{m_n^2-M^2}y),
	&
	f_{3R}\propto \cos (\sqrt{m_n^2-M^2}|y|+\alpha_3),
\end{cases}
\end{equation}
where the $\alpha$'s in the above argument  are defined as
\begin{equation}
	\begin{cases}
	\dis\cos\alpha_1=\frac{\sqrt{m_n^2-M^2}}{m_n}, \quad 
	\dis\sin\alpha_1=-\frac{M}{m_n},\\
	\dis\cos\alpha_\SM=-\frac{\sqrt{m_n^2-M^2}}{m_n}, \quad
	\dis\sin\alpha_\SM=-\frac{M}{m_n},\\
\dis\tan\alpha_\H=\frac{2M-\pi RM_\B^2}{2\sqrt{ m_n^2-M^2}},\\
\dis\tan\alpha_3=\frac{-2M-\pi RM_\B^2}{2\sqrt{ m_n^2-M^2}}.
\end{cases}
\end{equation}
The brane mass terms $M_\B$ are considered to come from the underlying theory,
 such as a Grand Unified Theory, it is therefore much larger than the compactification scale.
Then we take the limit $M_\B \to \infty$ and it reduces to $\alpha_\H=\alpha_3=-\pi/2$.

To obtain the practical B.C.s leading to the KK eigenstates,
we first calculate the phase matrix.
\begin{align}
\e^{i\frac{g_4 v}{2}\Sigma_1 y}
=
\begin{pmatrix}
(1,1)&(1,2)\\(2,1)&(2,2)
\end{pmatrix},
\end{align}
The submatrices become
\begin{align}
	(1,1)
	=&
	\left[
	\frac{1+\sigma_3}{2}+\frac{1-\sigma_3}{2} \cos^2\theta
\right]\cos\frac{g_4 vy}{2}
+ \frac{1-\sigma_3}{2} \sin^2\theta \cos\frac{g_4 vy}{\sqrt2}
	+i \sigma_1\cos\theta \sin\frac{g_4 vy}{2},
	\\
	(1,2)
	=&
	\sigma^- \sin\theta\cos\theta
	\left(\cos\frac{g_4 vy}{\sqrt2}-\cos\frac{g_4 vy}{2}\right) 
	-i \frac{1+\sigma_3}{2} \sin\theta\sin\frac{g_4 vy}{2} 
	+i\frac{1-\sigma_3}{2} \sin\theta\sin\frac{g_4 vy}{\sqrt2},
	\\
	(2,1)
	=&
	\sigma^+ \sin\theta\cos\theta\left(\cos\frac{g_4 vy}{\sqrt2}-\cos\frac{g_4 vy}{2}\right) 
	-i\frac{1+\sigma_3}{2} \sin\theta\sin\frac{g_4 vy}{2}
	+i\frac{1-\sigma_3}{2} \sin\theta\sin\frac{g_4 vy}{\sqrt2},
	\\
	(2,2)=&
	\frac{1+\sigma_3}{2} \sin^2\theta \cos\frac{g_4 vy}{2}
	+\left[\frac{1-\sigma_3}{2} + \frac{1+\sigma_3}{2} \cos^2\theta \right]\cos\frac{g_4 vy}{\sqrt{2}}
	+i \sigma_1\cos\theta\sin\frac{g_4 vy}{\sqrt2},
\end{align}
where $\sigma_{1,2,3}$ are  the Pauli matrices and $\sigma^\pm= \frac{\sigma_1\pm i\sigma_2}{2}$.
The B.C.s on the left-handed part from the continuous condition (\ref{summary_BC})
$\left.[\e^{i\frac{g_4 v}{2}y\Sigma_1}\hat {\bs \tau} (y)]_{\rm odd}\right|_{y=\pi R}=0$
 is given by extracting the odd function from each part.
\begin{align}
0
=&
\left.[\e^{i\frac{g_4 v}{2}y\Sigma_1}\hat {\bs \tau} (y)]_{\rm odd}\right|_{y=\pi R}
\nt
\\
\supset&
	\label{cond_KKmass_L1}
\begin{pmatrix}
	\cos\frac{\lambda}{2}f_{1L}& i\cos\theta\sin\frac{\lambda}{2}f_{\SM L}&-i\sin\theta\sin\frac{\lambda}{2}f_{\H L}&0
	\\
	0&i\sin\theta\sin\frac{\lambda}{\sqrt2}f_{\SM L} & i\cos\theta \sin\frac{\lambda}{\sqrt2}f_{\H L}&\cos\frac{\lambda}{\sqrt2}f_{3L}
\end{pmatrix}
\hat {\bs \tau}^{(n)}_{L},
\end{align}
where $\lambda =g_4 v\pi R$.
The $f_{1L/\SM L/\H L/3L}$ in the above expressions are understood to be the values of mode functions at $y=\pi R$.  

Next we discuss the conditions on the derivatives of mode function in eq. (\ref{summary_BC})
\begin{equation}
	\left.\left[\e^{i\frac{g_4 v}{2}\Sigma_1 y}\gamma_5\partial_y\hat {\bs \tau}(y)\right]_{\rm odd}\right|_{y=\pi R}
=
-M\left.\left[\e^{i\frac{g_4 v}{2}\Sigma_1 y}\hat {\bs\tau}(y)\right]_{\rm even}\right|_{y=\pi R}.
\end{equation}
The conditions on the derivatives  of mode functions are obtained 
 from the lower part of eq. (\ref{summary_BC}) as
\begin{align}
	\label{cond_KKmass_L2}
	&0=\nt
	\\
	&
	\left[\begin{array}{cc}
		i\cos\theta\sin\frac{\lambda}{2}(\partial_y+M) f_{1L}
		&
		(\cos^2\theta\cos\frac{\lambda}{2}+\sin^2\theta\cos\frac{\lambda}{\sqrt2})(\partial_y+M)f_{\SM L}
		\\
		-i\sin\theta\sin\frac{\lambda}{2}(\partial_y+M)f_{1L}
		&
		\sin\theta\cos\theta(\cos\frac{\lambda}{\sqrt2}-\cos\frac{\lambda}{2})(\partial_y+M)f_{\SM L}
	\end{array}\right. \nonumber
	\\
	&\hspace{25mm}
	\left.\begin{array}{cc}
		\sin\theta\cos\theta(\cos\frac{\lambda}{\sqrt2}-\cos\frac{\lambda}{2})(\partial_y+M)f_{\H L}
		&
		i\sin\theta\sin\frac{\lambda}{\sqrt2}(\partial_y+M)f_{3L}
		\\
		(\sin^2\theta\cos\frac{\lambda}{2}+\cos^2\theta\cos\frac{\lambda}{\sqrt2})(\partial_y+M)f_{\H L}
		&
		i\cos\theta\sin\frac{\lambda}{\sqrt2} (\partial_y+M)f_{3L}
	\end{array}\right]
	\hat {\bs \tau}^{(n)}_{L}. 
\end{align}
Note that the derivatives of the mode functions in the above are also understood to be $\partial_y f=\left. \partial_y f \right|_{y=\pi R}$.

Combining these two conditions (\ref{cond_KKmass_L1}) and (\ref{cond_KKmass_L2}), we have
\begin{align}
	0=&
	\left[
	\begin{matrix}
		\cos\frac{\lambda}{2}f_{1L} 
		&i\cos\theta\sin\frac{\lambda}{2} f_{\SM L}
		\\
		0
		&i\sin\theta\sin\frac{\lambda}{\sqrt2}f_{\SM L}
		\\
		i\cos\theta \sin\frac{\lambda}{2}(\partial_y+M)f_{1L}
		&(\cos^2\theta \cos\frac{\lambda}{2}+\sin^2\theta\cos\frac{\lambda}{\sqrt2})(\partial_y+M)f_{\SM L}
		\\
		-i\sin\theta\sin\frac{\lambda}{2}(\partial_y+M)f_{1L}
		&\sin\theta\cos\theta(\cos\frac{\lambda}{\sqrt2}-\cos\frac{\lambda}{2})(\partial_y+M)f_{\SM L}
	\end{matrix}
	\right. \nonumber 
	\\
	&\hspace{30mm}
	\left.
	\begin{matrix}
		-i\sin\theta \sin \frac{\lambda}{2}f_{\H L}
		&0
		\\
		i\cos\theta \sin\frac{\lambda}{\sqrt2}f_{\H L}
		&\cos\frac{\lambda}{\sqrt2}f_{3L}
		\\
		\sin\theta\cos\theta(\cos\frac{\lambda}{\sqrt2}-\cos\frac{\lambda}{2})(\partial_y+M)f_{\H L}
		&i\sin\theta\sin\frac{\lambda}{\sqrt2}(\partial_y+M)f_{3L}
		\\
		(\sin^2\theta\cos\frac{\lambda}{2}+\cos^2\theta\cos\frac{\lambda}{\sqrt2})(\partial_y+M)f_{\H L}
		&i\cos\theta\sin\frac{\lambda}{\sqrt2}(\partial_y+M)f_{3L}
	\end{matrix}
\right]\hat{\bs \tau}_L^{(n)}. 
\end{align}
To have non-trivial solutions of $\hat{\bs \tau}_L^{(n)}$,
the determinant of the above matrix must be vanished.
Substituting the mode functions (\ref{mode_function}) into the above matrix, the determinant takes the following form
\begin{align}
	0=&
	\left|
	\begin{matrix}
		\cos\frac{\lambda}{2} \sin\phi_n
		&i\cos\theta\sin\frac{\lambda}{2} \cos(\phi_n-\alpha_1)
		\\
		0
		&i\sin\theta\sin\frac{\lambda}{\sqrt2} \cos(\phi_n-\alpha_1)
		\\
		i\cos\theta \sin\frac{\lambda}{2}\cos(\phi_n+\alpha_1)
		&-(\cos^2\theta \cos\frac{\lambda}{2}+\sin^2\theta\cos\frac{\lambda}{\sqrt2}) \sin\phi_n
		\\
		-i\sin\theta\sin\frac{\lambda}{2}\cos(\phi_n+\alpha_1)
		&-\sin\theta\cos\theta(\cos\frac{\lambda}{\sqrt2}-\cos\frac{\lambda}{2})\sin\phi_n
	\end{matrix}
	\right.
	\nt
	\\
	&\hspace{20mm}
	\left.
	\begin{matrix}
	-i\sin\theta \sin \frac{\lambda}{2}\cos(\phi_n+\alpha_\H)
	&0
	\\
	i\cos\theta \sin\frac{\lambda}{\sqrt2}\cos(\phi_n+\alpha_\H)
	&\cos\frac{\lambda}{\sqrt2}\sin\phi_n
	\\
	-\sin\theta\cos\theta(\cos\frac{\lambda}{\sqrt2}-\cos\frac{\lambda}{2})\sin(\phi_n+\alpha_\H+\alpha_1)
	&i\sin\theta\sin\frac{\lambda}{\sqrt2}\cos(\phi_n+\alpha_1)
	\\
	-(\sin^2\theta\cos\frac{\lambda}{2}+\cos^2\theta\cos\frac{\lambda}{\sqrt2})\sin(\phi_n+\alpha_\H+\alpha_1)
	&i\cos\theta\sin\frac{\lambda}{\sqrt2}\cos(\phi_n+\alpha_1)
	\end{matrix}
	\right|
	\label{KK_mass_condition1}
	\\
	=&
	\cos(\phi_n+\alpha_1) \sin\phi_n
	\left[ 
		-\sin^2\phi_n+
		\left\{ 
			\sin^2\frac{\lambda}{2}+\left( \sin^2\frac{\lambda}{\sqrt2}-\sin^2\frac{\lambda}{2} \right)\sin^2\theta
		\right\}
		\cos^2\alpha_1
		\right],
	\end{align}
where we employ $\phi_n$ as $\pi R\sqrt{m_n^2 -M^2}$.
The phase $\alpha_\H$ is here set to be $-\frac{\pi}{2}$. 

Then, we find the three conditions of vanishing determinant as   
	\begin{equation}
		\label{KK_mass_condition}
	\begin{cases}
		\cos(\phi_n+\alpha_1)=0,\\
		\sin\phi_n=0,\\
		\frac{\sin^2\phi_n}{\cos^2\alpha_1}=\sin^2\frac{\lambda}{2}-\left( \sin^2\frac{\lambda}{2}-\sin^2\frac{\lambda}{\sqrt2} \right)\sin^2\theta
		=\sin^2\frac{\lambda}{2}\cos^2\theta+\sin^2\frac{\lambda}{\sqrt2}\sin^2\theta.
	\end{cases}
\end{equation}
The last condition depending on the Higgs VEV $\lambda=g_4v \pi R$ includes a zero-mode fermion
 since the SM fermions get a mass from the VEV of the Higgs field.
Note that the mixing parameter $\theta$ appears in the right hand side of the third line in (\ref{KK_mass_condition})
 due to the mixing of $\tau$ and $\tau'$. 
 When $\theta =0$,
 it is expected that the above result reduces to the case where only the representation $\bf 3$ is introduced. 
In fact, our obtained result agrees with the results in \cite{GHYukawaflat} as we expected.

\subsection{Deviation of yukawa couplings in GHU from the SM one}
In the above subsection, we have derived an analytic formula determining the KK mass spectrum (\ref{KK_mass_condition}).
The last formula in (\ref{KK_mass_condition}) provides the SM fermion spectrum.  
Although it cannot be analytically solved in terms of $m_n(v)$, we can still get the exact form of their derivatives:
$\frac{\d}{\d v} m_n(v)$.
Expressing the last formula in the following way
\begin{equation}
	\sin^2\phi_n = \left( 1-\frac{M^2}{m_n^2} \right)F(\lambda),
\end{equation}
where 
\begin{equation}
	F(\lambda)= \sin^2\frac{\lambda}{2} - \left( \sin^2\frac{\lambda}{2}-\sin^2\frac{\lambda}{\sqrt{2}} \right)\sin^2\theta,
\end{equation}
it is straightforward to calculate the derivative of $m_n(v)$,  
\begin{equation}
	\frac{\d m_n}{\d v}
	=
	\frac{g_4}{2}
	\frac{\phi_n\cos\alpha_1}{\phi_n\cot \phi_n-\sin^2\alpha_1}
	\frac{1}{F(\lambda)} \frac{\d F(\lambda)}{\d \lambda},
\end{equation}
where eq.~(\ref{KK_mass_condition}), $\cos\alpha_1=\frac{\sqrt{m_n^2-M^2}}{{m_n}}$ and $\sin\alpha_1=-\frac{M}{m_n}$ are used. 

This is yukawa coupling in GHU what we would like to obtain 
 and it varies according to the Higgs VEV $v$. 
To compare it with the SM yukawa coupling,
 we focus on the zero mode ($n=0$) sector and the ratio of them is found to be
\begin{equation}
\frac{f}{f_\SM} 
=
\frac{\frac{\d m_0}{\d v}}{\frac{m_0}{v}}
= 
\frac{\lambda}{2\pi R m_0} \frac{\phi_0 \cos\alpha_1}{\phi_0 \cot \phi_0 -\sin^2\alpha_1} \frac{1}{F(\lambda)} \frac{\d F(\lambda)}{\d \lambda}. 
\label{ratio}
\end{equation}
For the case $M>m_0$ (the zero mode is plausible),
 we should replace $\phi_0$ with $i\pi R \sqrt{M^2-m_0^2}$, and then,
 we have
\begin{equation}
	\label{Deviation_yukawa}
	\frac{f}{f_\SM}
	=\frac{\lambda}{2} \frac{M^2-m_0^2}{M^2-\pi RMm_0^2\sqrt{M^2-m_0^2}\coth (\pi R\sqrt{M^2-m_0^2})} 
	\frac{\d }{\d \lambda} \ln(F(\lambda)).
\end{equation}

\subsection{Numerical study}
\begin{figure}[h]
	\def\SCALE{1.6}
	\centering
	\includegraphics[scale=\SCALE]{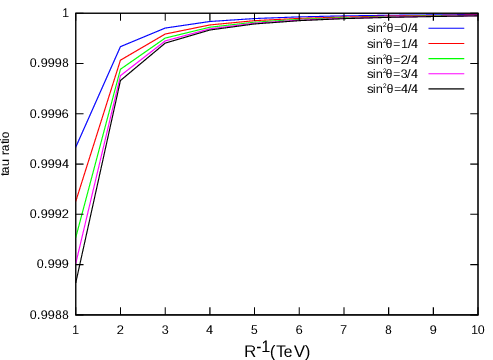}
	\includegraphics[scale=\SCALE]{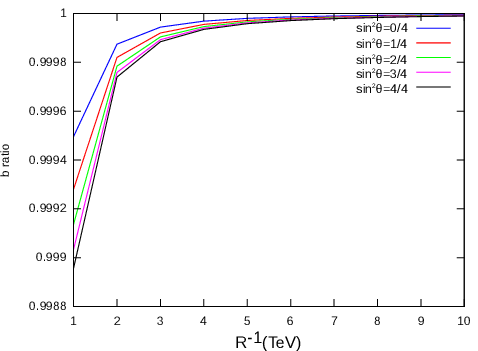}
	\caption{The plots of the ratio of the yukawa coupling as a function of the compactification scale are shown. 
	 The left (right) one is the case for the tau lepton (the bottom quark).}
	\label{fig:KKplot}
\end{figure}

In this subsection, we apply the above result (\ref{Deviation_yukawa}) to the tau lepton and the bottom quark 
 which would be measured at the LHC or ILC more promising than those of other fermions. 
Moreover, the tau lepton and the bottom quark can be assigned to 
 the same fundamental representation of $SU(3)$ in GHU as mentioned above, 
 therefore the result (\ref{Deviation_yukawa}) can be independently applied to the both cases. 
Regarding the fermion mass as an input parameter $m_0=m_\tau(m_b)$ 
 and rewriting the Higgs VEV by W-boson mass through $M_W=gv/2$, 
 the analytic formula determining the fermion mass and the ratio of the yukawa coupling are the following 
\begin{align}
	\label{KK_mass_condition2}
	&
	\dis
	\sinh^2\left[ \pi  R\sqrt{M^2-m_{\tau(b)}^2} \right]
	=
	\frac{M^2-m_{\tau(b)}^2}{m_{\tau(b)}^2} \nonumber 
	\\
	& 
	\hspace*{30mm}
	\times \left[ \sin^2\left(\pi RM_W \right)-\left( \sin^2\left( \pi RM_W \right)-\sin^2\left(\sqrt2 \pi RM_W \right) \right)\sin^2\theta \right],
	\\
	&
	\dis \frac{f}{f_\SM}
	=\dis
	\frac{M^2-m_{\tau(b)}^2}{M^2-\pi R m_{\tau(b)}^2\sqrt{M^2-m_{\tau(b)}^2}\coth(\pi R\sqrt{M^2-m_{\tau(b)}^2})}
	\nt	\\
	&
	\hspace{20mm}
	\times
	\dis
	\pi RM_W
	\frac{\sin\left( 2\pi RM_W \right) - 
		\left[\sin\left( 2\pi RM_W \right)-\sqrt2 \sin\left(2\sqrt{2}\pi RM_W  \right)\right]\sin^2\theta}
		{1-\cos(2\pi RM_W) - \left[\cos(2 \sqrt 2 \pi RM_W)-\cos\left( 2\pi RM_W \right)\right]\sin^2\theta}.
\end{align}
There are three parameters $R,M$ and  $\theta$ in our theory, 
 but one of them can be determined by the eq. (\ref{KK_mass_condition2}), that is to say, 
 the combination $RM$ is determined to reproduce the realistic fermion mass. 
We plot the ratio of the tau and the bottom yukawa coupling with some values of $\theta$ 
 as a function of the compactification scale $R^{-1}$ in Figure \ref{fig:KKplot}.
We see that the both yukawa coupling in GHU are almost consistent with the SM yukawa coupling, 
 which is also consistent with the present experimental data \cite{HiggsCoupling}.
It is quite natural since the weak scale $M_W$ is much smaller than the compactification scale. 
We also note that the periodicity of the VEV $v$ in yukawa coupling of GHU exists if and only if $\theta=0, \pi/2$. 
This is because yukawa coupling of GHU contains different two functions $\e^{2i\pi RM_W}$ and $\e^{2 \sqrt 2 i\pi RM_W}$ 
 with different periodicity in $v$.

\section{Summary}
In this paper, we have studied deviations of yukawa coupling in GHU from the SM one 
 by taking a five dimensional $SU(3)$ GHU model on the orbifold $S^1/Z_2$. 
It has been already pointed out in \cite{GHYukawaflat, GHYukawaWarped},
 the fermion mass $m(v)$ is periodic with respect to the Higgs VEV $v$ in GHU scenario.  
Then the derived yukawa coupling can be a nonlinear function of Higgs VEV $v$ with fermion bulk mass.
In the extreme case, the yukawa coupling vanishes at  $v=1/(g_4 R)$ 
 even though the fermions get nonzero mass from the Higgs VEV $v$. 
However, it is not clear that these properties are common in such scenario
 since several modifications are needed to construct realistic models. 

What should be emphasized in this paper is that 
 the brane mass terms and the additional different representations are taken into account 
 unlike the previous work \cite{GHYukawaflat}. 
In a realistic model of GHU, the brane mass terms are indispensable 
 not only to remove exotic massless fermions absent in the SM 
 but also to realize the flavor mixing as clarified in \cite{flavorGHU}. 
Since the brane mass terms change the boundary conditions for equation of motion and 
 this might change KK mass spectrum and yukawa coupling, 
 it is important for our study to incorporate the brane mass terms in realistic model of GHU.

We have derived an analytic formula determining KK mass spectrum and yukawa coupling. 
It can be shown that the nonlinear yukawa coupling still appears even in the case where the brane mass terms are considered. 
It should be noted that there is vanishing points of yukawa coupling in this theory 
 even though the periodicity of VEV $v$ is generically lost in the yukawa coupling 
 since it contains two different periodic function. 
This difference comes from the fact that the different kinds of representations are introduced 
 to construct more realistic models in GHU.   
As an application, we have numerically studied the ratio of tau and bottom yukawa couplings in GHU and the SM, 
 which would be measured at the LHC or ILC more promising than those of other lighter fermions. 
We have found that the both yukawa coupling in GHU are almost consistent with the SM yukawa coupling, 
 which is also consistent with the present experimental data \cite{HiggsCoupling}. 
It is expected that the consistency is true in any GHU scenario 
 since the Higgs VEV $v$ is much smaller than the compactification scale $R^{-1}$. 
We can always expand the yukawa couplings in terms of $vR$  
 and safely neglect the higher order terms of $vR$, 
{\em i.e.} the linear Higgs field approximation of yukawa coupling is always good picture in realistic parameter space.

\subsection*{Acknowledgments}
The work of N.M. is supported in part by the Grant-in-Aid 
 for Scientific Research from the Ministry of Education, 
 Science and Culture, Japan No. 24540283. 

\appendix
\section{Derivation of mode equations}
\label{Appendix1}
In this appendix, we describe the derivation of the mode equation, which is skipped in the main text.
Note that there are four kinds of fields, but their difference is whether the brane mass term exists or not. 
We first discuss the EOM of the $\hat \tau_1$ and $\hat\tau_\SM$ without the brane mass term.
Substituting mode expansions of $\hat\tau_1$ (\ref{mode_expansion}) into the EOM (\ref{EOM_tau2}), we have
\begin{align}
\begin{cases}
	i\partial_\mu\gamma^\mu \hat \tau_{1L}^{(n)}f_{1L}+(\partial_y-M\epsilon(y))\hat \tau_{1R}f_{1R}=0,
	\\
	i\partial_\mu\gamma^\mu \hat \tau_{1R}f_{1R}+(-\partial_y-M\epsilon(y))\hat \tau_{1L}f_{1L}=0.
\end{cases}
\end{align}
Replacing $i\partial_\mu\gamma^\mu\tau_{1L/R}^{(n)}$ with $m_n\tau_{1R/L}^{(n)}$,
the above mode equations become
\begin{align}
\begin{cases}
	m_{n} f_{1L}+(\partial_y-M\epsilon(y))f_{1R}=0,
	\\
	m_{n} f_{1R}+(-\partial_y-M\epsilon(y))f_{1L}=0.
\end{cases}
\end{align}
Since the $\tau_\SM$ obeys the same EOM of the $\tau_1$, 
 the corresponding mode functions $f_{\SM L}$ and $f_{\SM R}$ obey the same equations.

Next, we derive mode equations of $\hat \tau_\H,\hat \tau_3$.
To eliminate the brane localized fermion from the mode equation, 
 we multiply the conjugate of the differential operator of EOM from the left-hand side;
\begin{align}
&[-i\partial_\mu\gamma^\mu +\partial_y \gamma_5-M\epsilon(y)]
[i\partial_\mu\gamma^\mu -\partial_y \gamma_5-M\epsilon(y)]\hat \tau_\H 
 \nonumber \\
&=
-\sqrt{\pi R} M_B 
[-i\partial_\mu\gamma^\mu +\partial_y \gamma_5-M\epsilon(y)]
\tau_\B'\delta(y).
\end{align}
In the right hand side of the above equation, 
 ignoring the last term because of the property of the sign function: $\epsilon(0)=0$ 
 and the integration by parts of  the delta function $\int \d y f(y)\partial_y \delta(y)=-\int \d y \partial_y f(y) \delta(y)$,  
 we arrive at
\begin{align}
&[\partial^2-\partial_y^2-2M(\delta(y)-\delta(y-\pi R))\gamma_5+M^2]\hat\tau_\H=-\pi RM_B^2\hat \tau_\H\delta(y),
\end{align}
and the corresponding mode equation becomes
\begin{align}
\label{equation_H3}
\begin{cases}
	&[m_n^2-\partial_y^2-2M(\delta(y)-\delta(y-\pi R))+M^2]f_{\H L}=-\pi RM_B^2f_{\H L}\delta(y),
\\
&[m_n^2-\partial_y^2+2M(\delta(y)-\delta(y-\pi R))+M^2]f_{\H R}=-\pi RM_B^2f_{\H R}\delta(y).
\end{cases}
\end{align}
As mentioned at the beginning of this section, the mode functions $f_{3L}$ and $f_{3R}$ obey the same mode equations.

\section{Consistency checks of the analytic formula}
\label{Appendix2}
In this appendix, we check the consistency of the analytic formula determining KK mass spectrum in this model 
 by considering some specific cases. 
These observations support the validity of our analytic formula derived in this paper.  

\begin{itemize}

\item{\bf Case 1:} $M,M_\B\to 0$ and $\theta=0$

We first consider the simplest case vanishing the brane mass term and the mixing of $SU(2)$ doublets.
In this case, the analytic formula determining the KK mass spectrum is reduced to
\begin{align}
0
=
\sin\left( \phi_n+\frac{\lambda}{ 2} \right)
\sin\left( \phi_n-\frac{\lambda} {2} \right)
\sin\left( \phi_n+\frac{\lambda}{\sqrt 2} \right)
\sin\left( \phi_n-\frac{\lambda}{\sqrt 2} \right).
\end{align}
It indicates that the pattern of the mass spectrum is two kinds:
One is $\frac{n}{R} \pm \frac{g_4 v}{2}$ and the other is $\frac{n}{R}\pm\frac{g_4 v}{\sqrt2}$.
Due to the fact that the $\tau$ leptons in the $\bf 3$ and $\bf 6^\ast$ does not mix each other,
 their form of the KK mass spectrum retain the property of each representations.
Namely,
the $\tau$ leptons in the $\bf 3$ and $\bf 6^\ast$ have the yukawa coupling $\frac{g}{2}$
and $\frac{g}{\sqrt2}$ respectively.

\item {\bf Case 2:} $\theta \to 0, M_\B \to \infty$

Next, we discuss the case where the brane mass term exists. 
Since the mixing parameter $\theta $ is taken to be zero,
 the $\tau_\H$ is equivalent to $\tau'$ which come from the $\bf 6^\ast$.
Thus it is expected that the zero mode of $\tau'$  become massive by the brane mass $M_\B$.   
In this case, the analytic formula determining the KK mass spectrum is found as
\begin{align}
	0
	=\left[ \sin^2\phi_n-\sin^2\frac{\lambda}{2}\cos^2\alpha_1 \right]
	\sin\phi_n\cos(\phi_n-\alpha_1)
	\Rightarrow
	\begin{cases}
		\sin\phi_n=0,\\
		\cos(\phi_n-\alpha_1)=0,\\
		\sin^2\frac{\lambda}{2}=\frac{\sin^2\phi_n}{\cos^2\alpha_1},
	\end{cases}
	\label{}
\end{align}
where we set $\alpha_\H$ to be $-\pi/2$.
We notice that only the last condition depends on the Higgs VEV through $\lambda$ and includes the zero mode fermion mass.
For the case $m_n<M$, 
the last condition becomes
\begin{equation}
	\label{ex2_KK_mass}
\sin \frac{\lambda}{2}=\pm \frac{1}{\cos \alpha_1}\sin (\pi R\sqrt{m_n^2-M^2})
=\pm \frac{m_n}{i\sqrt{M^2-m_n^2}}i\sinh(\pi R\sqrt{M^2-m_n^2}),
\end{equation}
where we replace $\sqrt{m_n^2-M^2}$ with $i\sqrt{M^2-m_n^2}$. Then we have
\begin{equation}
\sin \frac{\lambda}{2}\simeq \pm \frac{m_n}{M}\sinh (\pi RM)
\Rightarrow m_n\simeq \pm \frac{M}{\sinh(\pi RM)}\sin\frac{\lambda}{2}. 
\end{equation}
Namely, the yukawa suppressions due to the bulk mass appears.
Moreover, the $\lambda\to 0$ recovers massless mode $m_n=0$.
This case reproduces the result in \cite{GHYukawaflat}. 

\item {\bf General case} $M_\B \to \infty$:
	
Finally, we consider the most general case, the $\bf 3$ and $\bf 6^\ast$ mix 
 in arbitrary angle $\theta$ which is discussed in the main text.
The mode functions at $y=\pi R$ in this case  become
\begin{align}
	f_{1L}\propto \sin\phi_n,
	f_{\SM L}\propto \cos(\phi_n-\alpha_1),
	f_{\H L}\propto \cos(\phi_n+\alpha_\H),
	f_{3L}\propto \sin\phi_n,
\end{align}
and the derivatives are given as
\begin{align}
	\begin{cases}
		(\partial_y+M)f_{1L}=m_n \cos(\phi_n+\alpha_1),\\
		(\partial_y+M)f_{\SM L}=-m_n \sin\phi_n,\\
		(\partial_y+M)f_{\H L}=-m_n \sin(\phi_n+\alpha_\H+\alpha_1),\\
		(\partial_y+M)f_{3 L}=m_n \cos(\phi_n+\alpha_1),
	\end{cases}
\end{align}
where we use $\cos\alpha_1=\sqrt{m_n^2-M^2}/m_n$ and $ \sin \alpha_1=-M/m_n$.
Then the KK mass condition will be 
\begin{align}
	0=&
	\left|
	\begin{matrix}
		\cos\frac{\lambda}{2} \sin\phi_n
		&i\cos\theta\sin\frac{\lambda}{2} \cos(\phi_n-\alpha_1)
		\\
		0
		&i\sin\theta\sin\frac{\lambda}{\sqrt2} \cos(\phi_n-\alpha_1)
		\\
		i\cos\theta \sin\frac{\lambda}{2}\cos(\phi_n+\alpha_1)
		&-(\cos^2\theta \cos\frac{\lambda}{2}+\sin^2\theta\cos\frac{\lambda}{\sqrt2}) \sin\phi_n
		\\
		-i\sin\theta\sin\frac{\lambda}{2}\cos(\phi_n+\alpha_1)
		&-\sin\theta\cos\theta(\cos\frac{\lambda}{\sqrt2}-\cos\frac{\lambda}{2})\sin\phi_n
	\end{matrix}
	\right.
	\\
	&\hspace{20mm}
	\left.
	\begin{matrix}
	-i\sin\theta \sin \frac{\lambda}{2}\cos(\phi_n+\alpha_\H)
	&0
	\\
	i\cos\theta \sin\frac{\lambda}{\sqrt2}\cos(\phi_n+\alpha_\H)
	&\cos\frac{\lambda}{\sqrt2}\sin\phi_n
	\\
	-\sin\theta\cos\theta(\cos\frac{\lambda}{\sqrt2}-\cos\frac{\lambda}{2})\sin(\phi_n+\alpha_\H+\alpha_1)
	&i\sin\theta\sin\frac{\lambda}{\sqrt2}\cos(\phi_n+\alpha_1)
	\\
	-(\sin^2\theta\cos\frac{\lambda}{2}+\cos^2\theta\cos\frac{\lambda}{\sqrt2})\sin(\phi_n+\alpha_\H+\alpha_1)
	&i\cos\theta\sin\frac{\lambda}{\sqrt2}\cos(\phi_n+\alpha_1)
	\end{matrix}
	\right|
	\\
	=&
	\cos(\phi_n+\alpha_1) \sin\phi_n
	\left[ 
		-\sin^2\phi_n+
		\left\{ 
			\sin^2\frac{\lambda}{2}+\left( \sin^2\frac{\lambda}{\sqrt2}-\sin^2\frac{\lambda}{2} \right)\sin^2\theta
		\right\}
		\cos^2\alpha_1
		\right]
	\\
	\Rightarrow&
	\begin{cases}
		\cos(\phi_n+\alpha_1)=0,\\
		\sin\phi_n=0,\\
		\frac{\sin^2\phi_n}{\cos^2\alpha_1}=\sin^2\frac{\lambda}{2}-\left( \sin^2\frac{\lambda}{2}-\sin^2\frac{\lambda}{\sqrt2} \right)\sin^2\theta
		=\sin^2\frac{\lambda}{2}\cos^2\theta+\sin^2\frac{\lambda}{\sqrt2}\sin^2\theta.
	\end{cases}
\end{align}
We can see that the zeromode mass $m_0$ which mass is provided by the VEV $v$ is corresponds to the third conditions
Then we have the KK mass conditions  as follows
\begin{align}
	\sin^2\frac{\lambda}{2} - 
	\left( \sin^2\frac{\lambda}{2}-\sin^2\frac{\lambda}{\sqrt2} \right)\sin^2\theta
	=&
	\frac{m_n^2}{m_n^2-M^2}\sin^2(\pi R\sqrt{m_n^2-M^2})
	\\
	=&
	\frac{m_n^2}{m_n^2-M^2}\sinh^2(\pi R\sqrt{M^2-m_n^2}).
\end{align}
The last expression corresponds to the case in $M>m_n$, especially to the zero mode.
The effects of mixture between the $\bf 3$ and $\bf 6^\ast$ reflects the 
left hand side of the above.
Namely,
if we set $\theta\to 0$, it recovers (\ref{ex2_KK_mass}).
On the other hand,
if we set $\theta \to \pi/2$ , the $\tau_\SM$ is equivalent to the $\tau$ in the $\bf 6^\ast$
so that the yukawa coupling is give by $\frac{g}{\sqrt2}$,
\begin{align}
	\sin^2\frac{\lambda}{\sqrt2}  
	=
	\frac{m_n^2}{m_n^2-M^2}\sinh^2(\pi R\sqrt{M^2-m_n^2}).
\end{align}
It corresponds to replace the $g/2$ in eq. (\ref{ex2_KK_mass}) with $g/\sqrt2$.

\end{itemize}


\end{document}